\documentclass[a4paper,11pt]{article}
\pdfoutput=1

\usepackage{jinstpub}

\usepackage{lineno}
\usepackage{epstopdf}
\usepackage{relsize}

\def\cesrta{{C{\smaller[2]ESR}TA}}

\title{Beam Position Monitoring System at CESR}

\author{M.G.Billing, W.F.Bergan, M.J.Forster, R.E.Meller, M.C.Rendina, N.T.Rider, D.C.Sagan, J.Shanks,
J.P.Sikora, M.G.Stedinger, C.R.Strohman,\
Cornell Laboratory for Accelerator-based ScienceS and Education (CLASSE),\
Cornell University,\ 161 Synchrotron Dr., Ithaca, NY, 14850, U.S.A.\\}
\author{M.A.Palmer,\
Collider Accelerator Department, Brookhaven National Laboratory,\
P.O.Box 5000, Upton, NY, 11973-5000, U.S.A.\\}
\author{R.L.Holtzapple,\
California Polytechnic State University, Physics Department,\
1 Grand Ave, San Luis Obispo, CA,  93407, U.S.A.}

\abstract{The Cornell Electron-positron Storage Ring (CESR) has been
converted from a High Energy Physics electron-positron collider to
operate as a dedicated synchrotron light source for the Cornell High
Energy Synchrotron Source (CHESS) and to conduct accelerator physics
research as a test accelerator, capable of studying topics relevant
to future damping rings, colliders and light sources.  Some of the
specific topics that were targeted for the initial phase of
operation of the storage ring in this mode, labeled \cesrta~(CESR as a Test Accelerator),
included 1)~tuning techniques to produce low emittance beams, 2)~the
study of electron cloud development in a storage ring and
3)~intra-beam scattering effects.  The complete conversion of CESR
to \cesrta~occurred over a several year period and is described
elsewhere\cite{JINST10:P07012}, \cite{JINST10:P07013},
\cite{JINST11:P04025}.  As a part of this conversion the CESR beam
position monitoring (CBPM) system was completely upgraded to provide
the needed instrumental capabilities for these studies.  This paper
describes the new CBPM system hardware, its function and
representative measurements performed by the upgraded system.}


\begin{document}
\maketitle
\flushbottom

%-----------------------------------------------%
% CesrTA Phase I Report                         %
% Chapter - CESR Conversion                     %
% Section - Beam Instrumentation and Feedback   %
% Section coordinator:  Mike Billing            %
% Page estimate:  24pp.                         %
%-----------------------------------------------%

\section[BPM Design]{Instrument Requirements and Design}
%\label{sec:cesr_conversion.beam_instr}

The \cesrta\ project\cite{JINST10:P07012, JINST10:P07013,
JINST11:P04025} required the development or upgrading of
several systems involved in accelerator operations. In particular, a
significant upgrade was needed for the beam position monitor (BPM)
system, which replaced an original relay-based position monitor
system.  The new individual readout modules for each BPM are capable
of turn-by-turn and bunch-by-bunch trajectory measurements for
bunches spaced as closely as 4~nsec or counter-rotating bunches with
a 14~nsec spacing.

The upgraded readout for the BPM system that provides high resolution
measurement capability has been designed and deployed.  This system
provides turn-by-turn measurements of individual bunches within bunch
trains with spacings that are multiples of either 4~nsec or 14~nsec.
The system has the ability to measure betatron phase, and coupling 
via synchronous detection of a driven beam.  This paper is the concluding 
document in a sequence preliminary and supplementary papers and reports on the 
CBPM system in publications\cite{PRSTAB8:062802, PRSTAB13:092802, PRSTAB14:072804} at 
accelerator conferences\cite{PAC01:TPAH064, PAC03:WPAG004, PAC03:WPPB027, PAC05:TPPP011, IPAC10:MOPE089, IPAC11:TUPC052}
and within the report on progress for the \cesrta~Project in the \cesrta~Phase~1 
Report\cite{CLNS:12:2084}.  Substantial sections of the paper for the
accelerator conference proceedings for IPAC~10\cite{IPAC10:MOPE089} have
been reproduced here verbatim.

\subsection{System Requirements}
%\label{sssec:cesr_conversion.beam_instr.bpm.sys_req}
The primary operational requirements for the CBPM system include:
\begin{itemize}
\item The ability to operate with counter-rotating beams of electrons
and positrons in a single vacuum chamber for two-beam synchrotron
light operations for CHESS;
\item High resolution for low emittance optics correction and tuning;
\item Turn-by-turn readout capability for multiple bunches to support
beam dynamics studies;
\item Capability for digitizing single species bunch trains with
bunch spacing as small as 4~nsec and dual beam digitization for bunch
trains with 14~nsec spacing.
\end{itemize}

The need for dual beam operation of the system places a unique constraint
on the CESR BPM specifications.  Since the relative arrival time of the bunches
from the two beams varies widely from location to location around the ring,
standard RF processing techniques to optimize resolution and minimize timing
sensitivity cannot be applied to the full system. As a result, the CESR design
utilizes peak sampling with a high bandwidth digitizer and incorporates hardware
and software design features to optimize the system timing performance.
Table ~\ref{tab:BPMRequirements} summarizes the design specifications for the
high resolution measurements required for low emittance optics correction
\cite{IPAC10:MOPE089}.

\begin{table}[h]
    \centering
    \caption{\label{tab:BPMRequirements} CESR BPM Module Requirements\cite{IPAC10:MOPE089}.}
    \vspace*{1 ex}
    \begin{tabular}{ll}
        \hline\hline
        Parameter & Specification\\
        \hline
        Front End Bandwidth (required for 4 ns bunch trains) & 500 MHz \\
        Absolute Position Accuracy (long term) & 100 $\mu$m \\
        Single Shot Position Resolution & 10 $\mu$m \\
        Differential Position Accuracy & 10 $\mu$m \\
        Channel-To-Channel Sampling Time Resolution & 10 psec \\
        \hline\hline
    \end{tabular}
\end{table}

\subsection{System Design}
\begin{figure}[h] %  figure placement: here, top, bottom, or page
   \centering
   \includegraphics[height=4in]{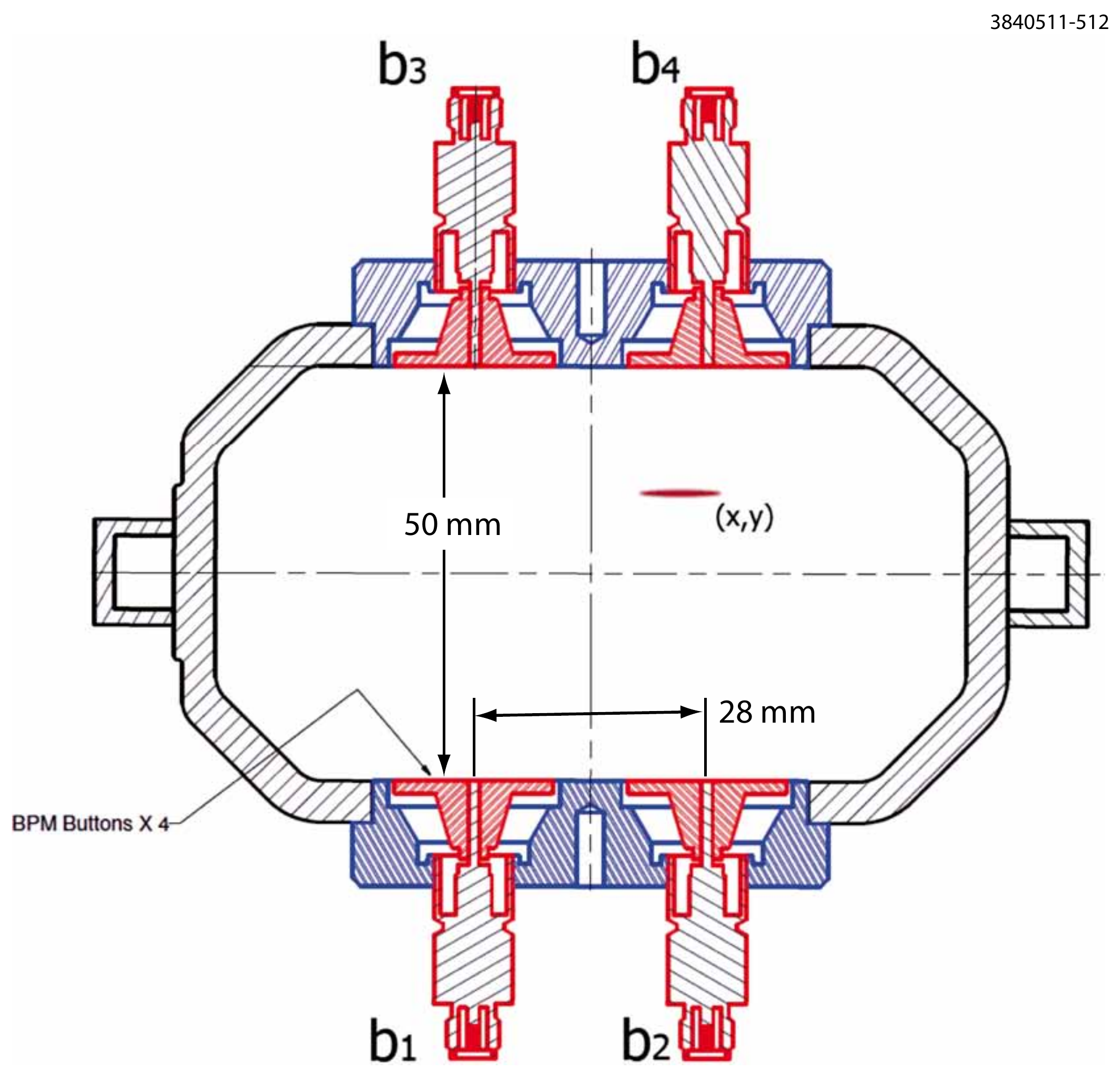}
   \caption[CESR Beam Position Monitor]{\label{fig:CESRBPM} BPM detector as
   configured in one of the superconducting wiggler vacuum chambers.}
\end{figure}

%\label{sssec:cesr_conversion.beam_instr.bpm.sys_des}
The CESR BPM system consists of a network of local sensors and processors.
Each location has four beam buttons arranged in a mirror symmetric fashion,
an example of which is shown in Figure~\ref{fig:CESRBPM}, providing signals
for each processing module.  The relative amplitude of the four BPM electrodes
yields horizontal and vertical position and beam intensity information.
All modules share a common control database, timing and synchronization
controls, and networked data storage.  This allows for accelerator-wide
coordinated measurements.

\begin{figure}[h] %  figure placement: here, top, bottom, or page
   \centering
   \includegraphics[height=4in]{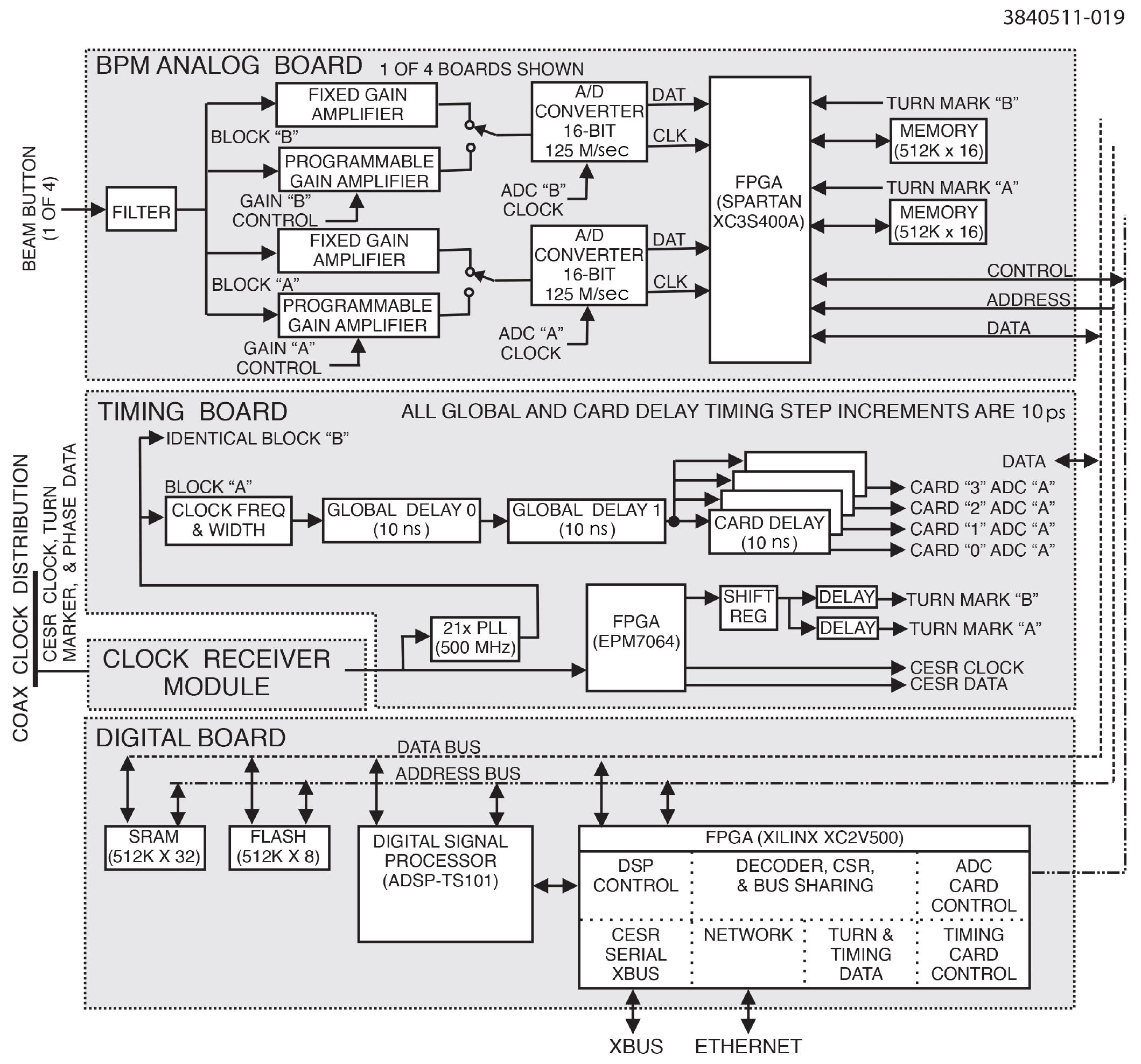}
   \caption[BPM module functional diagram.]{\label{fig:BPM Module Functional}
   BPM module functional diagram\cite{IPAC10:MOPE089}.}
\end{figure}

Figure~\ref{fig:BPM Module Functional} shows a block diagram of the digital
BPM readout modules developed for CESR. Each module incorporates four front
end boards each with two parallel 16-bit digitizer chains based on the Analog
Devices AD9461 operating with digitization rates up to 125~MHz. When operating
with 4~ns bunch trains, digitizing is interleaved between the two chains. For
14~ns dual species operation, each digitizer chain handles a single species.
The front end boards have 1)~a fixed gain amplifier optimized for precision
measurements for bunches with $ N_{b} \sim 1\times10^{10}$ particles and
2)~a digitally-controlled variable gain amplifier permitting measurements
over a wide dynamic range for the beam current.

\begin{figure}[h] %  figure placement: here, top, bottom, or page
   \centering
   \includegraphics[height =2in]{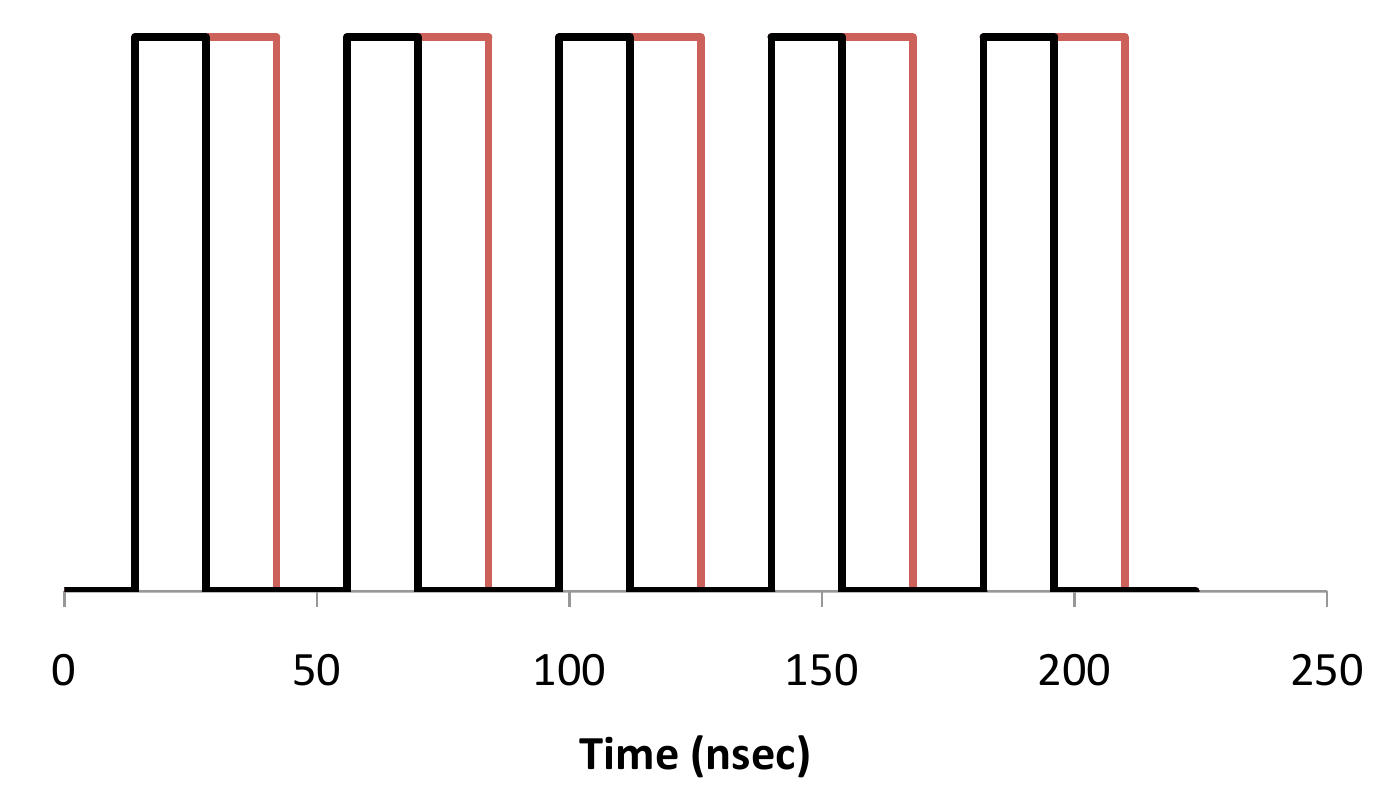}
   \caption[CESR Beam Position Monitor Timing Signal]{\label{fig:TimingSignal}
   Example of a CBPM 24~MHz timing signal.  The black portion of the waveform is the
   same for all periods of the timing signal.  The red portion of the waveform may be
   a digital high or low in order to carry a single bit of information for that clock
   period. In the next clock period the digital sense of the red signal is reversed in
   order to prevent a DC shift in the waveform, which would  depend on the information
   being carried in that third timing state.  (A DC shift of the waveform could add
   jitter to the timing edges.) As the there are 61 periods of the 24~MHz signal per
   CESR turn, the trigger signal actually encodes 28 bits of information, which is sent
   to all modules per turn.}
\end{figure}

The timing configuration is implemented by a dedicated timing board
integral within each module, which takes as an input a
three-temporal-state 24~MHz clock reference signal from the CESR
master timing system. As seen in Figure~\ref{fig:TimingSignal} the
rising clock edges are phase-locked to a 24~MHz clock on the timing
board.  The second timing slot per 24~MHz period encodes 1) a turns
marker signal, a signal selecting the turn for all CBPM modules to
initiate data acquisition, 2) two hardware triggers (to initiate
synchronous data acquisition), 3) 8 bits for a software command
word, and 4) the 9-bit phases for each of the two phase lock
loops from the (horizontal, and vertical) tune trackers.  To avoid a
timing slew as the bit pattern changes, the data is encoded into
every other 24~MHz period and its complement is encoded into the
immediately following 24~MHz period. The turns marker signal is not
encoded as ordinary data, instead it is encoded as a "code
violation". For the turns marker, the sequencing of bit and
complement of that bit is violated by sending the same state in 3
consecutive bits. This could never happen with ordinary data. There
is also a guard bit that follows normal complementing at each end of
this 3-bit stream, hence the use of 5 of the 61 bits to encode the
turns marker.  The timing board provides overall digitization rate
control, adjustment capability for channel-to-channel digitization
time delays, and global adjustment capability for the module
digitization time relative to the bunch arrival time at the
detector. The timing delays have a resolution of $\Delta
t_{\textrm{step}} = 10$~psec and this degree of local timing
adjustment is required to sample at the peak of each BPM signal in
order to maintain the resolution and noise performance for each
device.

Communications, operational control, and onboard data processing for each device is
provided through a digital board and TigerSharc digital signal processor (DSP).
Communication is supported for both ethernet and the dedicated CESR field bus, 
Xbus\cite{IEEENS26:4078to4079, IPAC12:THPPR015}.

The CESR BPM system is controlled via custom server applications running on the CESR
control system cluster.  All system parameters, including pedestal and gain scale calibration
values, delay tables for all supported bunch spacing configurations, and various
identification and management data, are stored in a central location.  Each server application
is responsible for loading, managing, and saving this information with new values generated
within the instruments.  All control and data read back is performed via a custom network
protocol\cite{Strohman:CBINET} running over 100~Mbit ethernet.

Rapid measurements, taking advantage of on-board averaging, pedestal subtraction, and gain
scaling can be requested of all or a subset of instruments or detailed turn-by-turn data can
be acquired for an arbitrary combination of bunches and turns in 14~ns or 4~ns bunch spacing
modes.  The software on board the instruments is also capable of automatically determining
appropriate delays to use for sampling at the optimal point of the incoming waveform on all
channels.

Data from all detectors are stored in a centrally-located database.
Raw ADC values, along with pedestals and gain scale factors for all
channels and amplifier settings are stored in all data files.
Sufficient information is provided to allow analysis of raw ADC or
pedestal-subtracted and gain-scaled turn-by-turn button data and/or
physical beam positions for every bunch stored and detector location
in the machine.

\section{Measurements and Characterization}

\subsection{Acquisition of Position and Trajectory Data}

Position data may be acquired in two basic modes: orbit data and
turn-by-turn trajectory data. For turn-by-turn measurements the acquisition in all
CBPM modules is initiated by a trigger bit appearing in the timing
signal.  This bit is typically timed to occur several CESR periods
(2.56~$\mu$sec) before the turn during which beam transfer would
occur from the Synchrotron injector into CESR. This permits the
measuring of injected beam into CESR, when there is no stored beam.
Since the transfer trigger is synchronous with the 60~Hz mains
frequency $\pm$75~$\mu$sec (the delay to align the the timing of 
bunches in the Synchrotron injector and CESR) and the number of turns may be chosen to
span an integer number of mains frequency cycles, any averaging over
such a span would produce the same average when performed for
different measurements.

\begin{figure}[h] %  figure placement: here, top, bottom, or page
   \centering
   \includegraphics[height=3in]{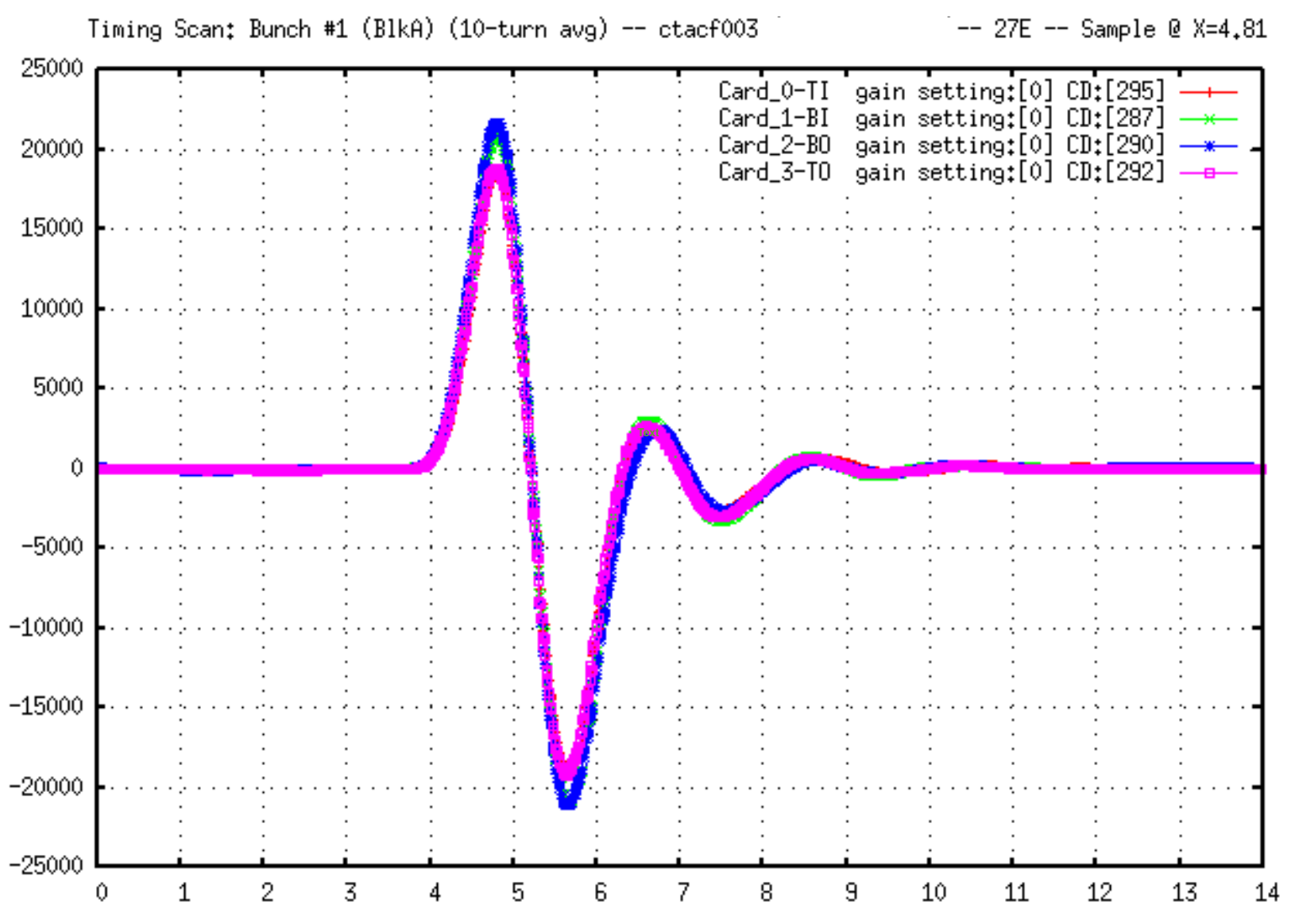}
   \caption[Timing scan of the four BPM electrodes.]{\label{fig:BPM_SIgnalSampling}
   Timing scan of the four BPM electrodes from BPM27E with the beam nearly on-axis using the
   fixed gain channel with a positron bunch of 0.75 mA.  The vertical scale is digitizer units
   and the horizontal scale is nsec.  Each point of the scan is a ten turn average.  The
   difference in the effective period of the oscillation present in these filtered signals
   is due to the tolerances of the filter components.}
\end{figure}

The incoming signals are filtered to produce a waveform with a
longer time shape to reduce the sensitivity to the sampling time.
Since the signal coming from the button electrode is essentially a
differentiated gaussian, after filtering, the signal shape for a
positron bunch becomes a 500~MHz sine-wave. The initial end of the
500~MHz signal connects smoothly to the DC baseline with an
approximately constant curvature arc.
Figure~\ref{fig:BPM_SIgnalSampling} shows a sampling time scan for
the four electrode signals at one CBPM modules.  After filtering,
the sinusoidal portion of the waveform shape for the button signal
from the n-th electrode, $b_n(t)$, may be written as a function of
time in terms of the peak signal $\hat b_n$ as

\begin{eqnarray}
b_n(t) = \hat b_n \cos \{2\pi f( t + \Delta t_n )\}
\end{eqnarray}

\noindent where n ranges from 1 to 4, $\Delta t_n$ is the average temporal offset of the
sampling gate with respect to the first peak of the waveform and $f$
is approximately 500~MHz.

During orbit measurements typically the four electrode signals are acquired for 1024 turns.
For small displacements each electrode's signal is averaged and the average x- and y-positions
are computed using

\begin{eqnarray}
x &=& k_x\frac{\Delta_x}{\Sigma}\nonumber\\
y &=& k_y\frac{\Delta_y}{\Sigma}\label{eqn:bi_to_xy}
\end{eqnarray}

\noindent where $\Delta_x$ is the difference of the radial outside
electrodes minus the radial inside electrodes, $\Delta_y$ is the
difference of the  upper electrodes minus the lower electrodes and
$\Sigma$ is the sum of all four electrodes.  The vacuum
chamber-specific geometry factors, $k_x$ and $k_y$, are given in
Table~\ref{tab:BPMGeometricFactors} for the different types of
vacuum chambers in CESR.

\begin{table}[h]
    \centering
    \caption{\label{tab:BPMGeometricFactors} Geometric Factors for Different CESR Vacuum Chamber Cross Sections}
    \vspace*{1 ex}
    \begin{tabular}{ l c c}
        \hline\hline
        Vacuum Chamber Geometry & $k_x$ (mm) & $k_y$ (mm)\\
        \hline
        Normal CESR Cross Section (50 mm x 90 mm)\cite{PRSTAB8:062802} & 26.2 & 19.6\\
        CESR Round Cross Section (45.0 mm inner radius)\cite{BIW92:55to77} & 22.5 & 22.5\\
        Undulator Cross Section (5 mm x 70 mm)\cite{Billing:UNDBPM} & 2.43 & 1.83\\
        \hline\hline
    \end{tabular}
\end{table}

%k_x = 26.25e-3 + 26.04x**2 - 2551 x**4  \  CESR
%k_y = 19.63e-3 + 12.63y**2 + 152y**4    /  Normal

For turn-by-turn trajectories the number of turns may be chosen in the calling sequence to
the CBPM servers.  One of the trigger bits within the timing signal starts the turn-by-turn
digitization of the electrode signals.  After acquiring the complete set of data, the positions
are computed turn by turn using the same formulae as the orbit data.  The entire position sequence
is returned and stored in a raw data file.

\subsection{Estimating Position Uncertainties}

The uncertainties in position measurements $\delta x, \delta y$ may
be directly attributed to the uncertainty in individual button
signals $\delta b_n/b_n$. Starting from Eqn. \ref{eqn:bi_to_xy}:

\begin{eqnarray}
\delta x &=& \sqrt{\sum_n {\left(\frac{\partial x}{\partial
b_n}\delta b_n\right)^2}},\qquad \textrm{where} \nonumber\\
\left|\frac{\partial x}{\partial b_n}\right| &=& k_x
\left|\frac{1}{\Sigma} \pm \frac{\Delta_x}{\Sigma^2}\right|
\end{eqnarray}

\noindent If the beam is well-centered in the vacuum chamber, all
four button amplitudes $b_n$ will be comparable, and their
uncertainties $\delta b_n$ will also be comparable. Thus, $\Delta_x
\approx 0$ and $\left|\frac{\partial x}{\partial b_n} \right|\approx
\left|\frac{k_x}{\Sigma}\right| \approx \frac{1}{4}\left|\frac{k_x}{b_n}\right|$, and

\begin{eqnarray}
\delta x &\cong& \sqrt{4\left(\frac{\partial x}{\partial
b_n}\delta b_n\right)^2}\nonumber\\
&\cong& \frac{1}{2} k_x \left|\frac{\delta b_n}{\hat b_n}\right|
\end{eqnarray}

\noindent The error propagation for $\delta y$ yields a similar
result:

\begin{eqnarray}
\delta y &\cong& \frac{1}{2} k_y \left|\frac{\delta b_n}{\hat
b_n}\right|
\end{eqnarray}

For BPM systems, which integrate over multiple bunch passages, noise
from the analog-to-digital converters (ADCs) dominates $\delta
b_n/\hat b_n$. For the CBPM fixed-gain processing channel, the noise
level observed in the 16-bit digitizer is two least significant bits
(LSB).  If the electrode signals are approximately 75\% of full
scale to allow for off-axis trajectories, this implies that the
amplitude measurement error $\delta \hat b_n$ due to digitizer
fractional sampling error is

\begin{eqnarray}
\frac{\delta \hat b_n}{\hat b_n} &\simeq& (0.75)^{-1} \times
2^{-14}\nonumber\\
 &=&8.1 \times 10^{-5}\label{eqn:dbn_dac}
\end{eqnarray}

For peak-detection BPMs with bunch-by-bunch resolution such as the
CBPM system, one must also account for timing errors. Therefore, the
single-turn position measurement plus its error should be estimated
by including not only an amplitude measurement error $\delta \hat
b_n$ due to ADC noise, but also the timing offset of the gate
$\Delta t_n$ and sampling trigger jitter error $\delta t$. Thus, the
measurement $b_n(t)$ plus its uncertainty $\delta b_n(t)$ are:

\begin{eqnarray}
b_n(t) + \delta b_n(t) = \left(\hat b_n + \delta \hat b_n \right)
\cos \left\{2\pi f( {t + \delta t + \Delta {t_n}}) \right\}
\end{eqnarray}

\noindent If the timing jitter is small, i.e., $2\pi f \delta t \ll 1$, then
to second order in $\delta t$ the amplitude measurement error may be
approximated as:

\begin{eqnarray}
\delta b_n(t) \cong -2\pi^2 f^2 \hat b_n \cos \{ 2 \pi f \Delta
{t_n} \}\delta t^2 - 2\pi  f \hat b_n\sin \{ 2\pi f \Delta
t_n\}\delta t + \delta \hat b_n \cos \{ 2 \pi f\Delta t_n
\}\label{eqn:dbn_exact}
\end{eqnarray}

\noindent From this expression, is clear that if the average timing offset
$\Delta t_n$ were exactly zero, the amplitude error would be
strictly second order in $\delta t$, with no linear component.
Additionally, this expression implies that, if all four electrode
signals had the same average timing offset $\Delta t_n$, the error
due to timing jitter $\delta t$ would cause all four signals to
scale together in amplitude, and would therefore not significantly
affect the position measurements. However, as long as the relative
timing offsets between the four electrode signals are different, the
signal amplitude uncertainties will have a first-order dependence on
the timing jitter $\delta t$.

If, in addition to the timing jitter $\delta t$ being small, the
gate timing error is also small (i.e., $2\pi f \Delta t_n \ll 1$),
this simplifies further to:

\begin{eqnarray}
\delta b_n(t) \cong -4\pi^2 f^2 \hat b_n \Delta t_n \delta t +
\delta \hat b_n \label{eqn:dbn_approx}
\end{eqnarray}

\noindent Note that Eqn. \ref{eqn:dbn_approx} explicitly shows a first order
dependance on $\delta t$.

As stated previously, the minimum timing delay step for each channel
is $\Delta t_{\textrm{step}}=10$~psec, corresponding to $\Delta
t_{\textrm{min}} = \pm 5$~psec precision. The time-in process is
independent for the four button channels on a given module,
therefore the typical timing offsets between different digitizing
channels is taken to be $\Delta t_n = \sqrt{2}\times \Delta
t_{\textrm{min}} = 7$~psec.

The typical timing jitter $\delta t$ can be directly measured using
the CBPM modules themselves. To do this, the sampling point is
shifted to near the zero-crossing of the 500~MHz filtered BPM
electrode signal shown in Fig. \ref{fig:BPM_SIgnalSampling}. At the
zero-crossing, the relative amplitude error scales proportionally
with the timing jitter $\delta t$:

\begin{eqnarray}
\delta b_{n,\,\textrm{timing}}(t) &\cong& \frac{\partial b_n}{\partial t}(t) \delta t \nonumber\\
&=& - 2\pi f \hat b_n \sin \{ 2\pi f ( t + \Delta t_n ) \} \delta t \nonumber\\
&\cong&  - 2 \pi f \hat b_n \delta t
\end{eqnarray}

\noindent where the waveform is taken to be zero at the location of
the greatest slope (i.e., no DC offset).

Turn-by-turn data was acquired for six BPM processors in 15
sequential sets of measurements of 16,384 turns, timed in at the
zero-crossing to maximize sensitivity to timing jitter. Using this
technique, the average timing jitter has been determined to be
$\delta t = 10.5$~psec. It is interesting to note if this data is
analyzed in 64-turn blocks (corresponding to roughly three
synchrotron oscillation periods), the RMS is $\delta t = 1.1$~psec, roughly
corresponding to the expected $1/\sqrt{N}$ for random noise.

These timing errors will produce an uncorrelated variation in the
button signals. Following Eqn. \ref{eqn:dbn_approx}, the
contribution from the gate timing $\Delta t_n$ and timing jitter
$\delta t$ to the fractional amplitude error signal is:

\begin{eqnarray}
\left| \frac{\delta b_n}{\hat b_n} \right|_{\textrm{timing}}
&\cong& 4 \pi^2 f^2 \Delta t_n \delta t \nonumber\\
&=& 4 \pi ^2 (500~\textrm{MHz})^2 (7~\textrm{psec})
(10.5~\textrm{psec}) \nonumber\\
&=& 7.25 \times 10^{-4}\label{eqn:dbn_timing}
\end{eqnarray}

The contributions from digitizing noise (Eqn. \ref{eqn:dbn_dac}) and
from timing jitter (Eqn. \ref{eqn:dbn_timing}) to the relative
signal amplitude are assumed to be uncorrelated, hence their
contributions add in quadrature. This gives a net fractional
uncertainty of $\delta b_n/\hat b_n = 7.29 \times 10^{-4}$ for each
of the electrode signals.

As a cross-check one can also directly measure the peak signal
variation. Turn-by-turn data was again acquired for six BPM
processors in 15 sequential sets of measurements of 16,384 turns,
timed in at the peak of the 500~MHz waveform. This yields an average
observed fractional variation of the peak button signal of $3.7
\times 10^{-3}$. This is a considerably larger variation compared to
the calculated fractional uncertainty of $\delta b_n/\hat b_n =
7.29\times 10^{-4}$. It is important to note that the direct
measurement of the peak signal variation also includes true beam
motion, which cannot be deconvolved from timing or ADC
contributions.

If the button signals are now averaged in 64-turn blocks, the
averaged fractional amplitude RMS becomes $8.7 \times 10^{-4}$,
roughly comparable to the computed $\frac{\delta b_n}{\hat b_n}$,
implying the button variations on timescales $> 64$~turns are due to
real beam motion. The fractional button signal error calculated from
the pair-wise BPM button differences for the 64-turns averages is
observed to be $5.5 \times 10^{-4}$, smaller than the expected
$\sqrt{2}\times (8.7\times 10^{-4}) = 1.2 \times 10^{-3}$ RMS, if
the noise were completely random. This implies that the button
signals themselves are varying systematically on a longer timescale
$> 64$~turns, attributed to variation in the timing jitter $\delta
t$ for the four button signals. If the gate delays $\Delta t_n$ were
accurately determined to place the sampling point at the peak of all
button signals, but the timing jitter $\delta t$ for all buttons
were moving back and forth over the quadratic peak of the signals in
a synchronous manner, the individual button signals would show a
variation. However, this variation would cancel out when examining
the pair-wise button differences. Since the beam position is
calculated from button differences, this effect would not impact
position measurements. Therefore, if the measured peak amplitude
variation is due to correlated timing deviations between the button
signals, much of the timing variation will have no influence on the
accuracy of the position measurements.

Therefore, the estimated position uncertainties for the standard
CESR vacuum chamber cross section are

\begin{eqnarray}
\delta x = \frac{1}{2} k_x \frac{\delta b_n}{\hat b_n} =
\frac{1}{2} (28~\textrm{mm})(7.3 \times 10^{-4}) &=& 10.2~\mu\textrm{m}\\
\delta y = \frac{1}{2} k_y \frac{\delta b_n}{\hat b_n} = \frac{1}{2}
(25~\textrm{mm})(7.3 \times 10^{-4}) &=& 9.1~\mu\textrm{m}
\end{eqnarray}

These estimates predict reasonable success for achieving the design
goals.  They may also be over-estimates, since it is assumed that
the beam centroid was not moving during the preceding measurements.

\begin{figure}[h]
\centering
\includegraphics[scale=0.32]{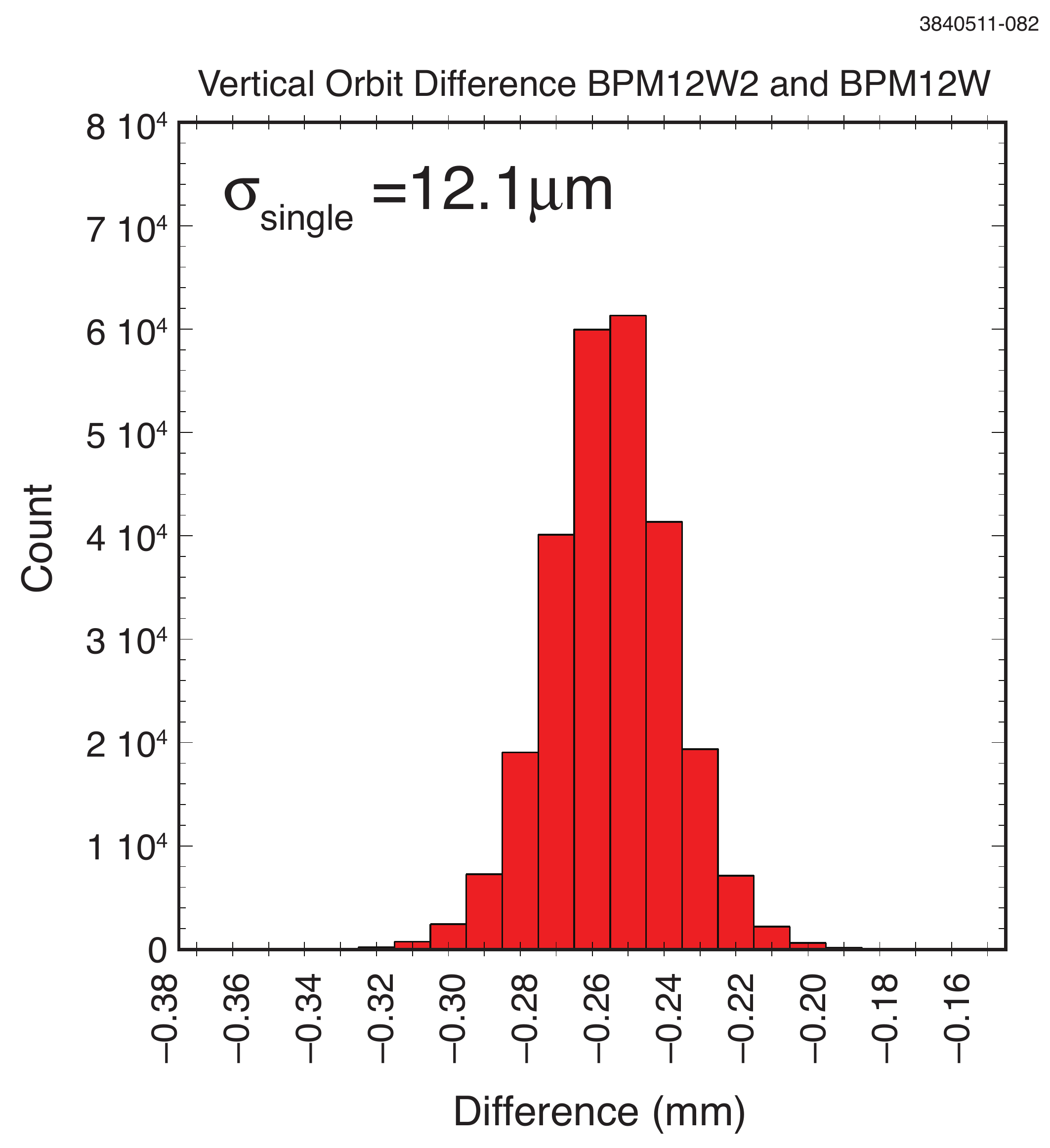}
%\caption[Vertical orbit differences between closely located detectors I.]{\label{fig:Triplet1} Vertical orbit differences between closely located detectors}
%end{figure}
%\centering
\includegraphics[scale=0.32]{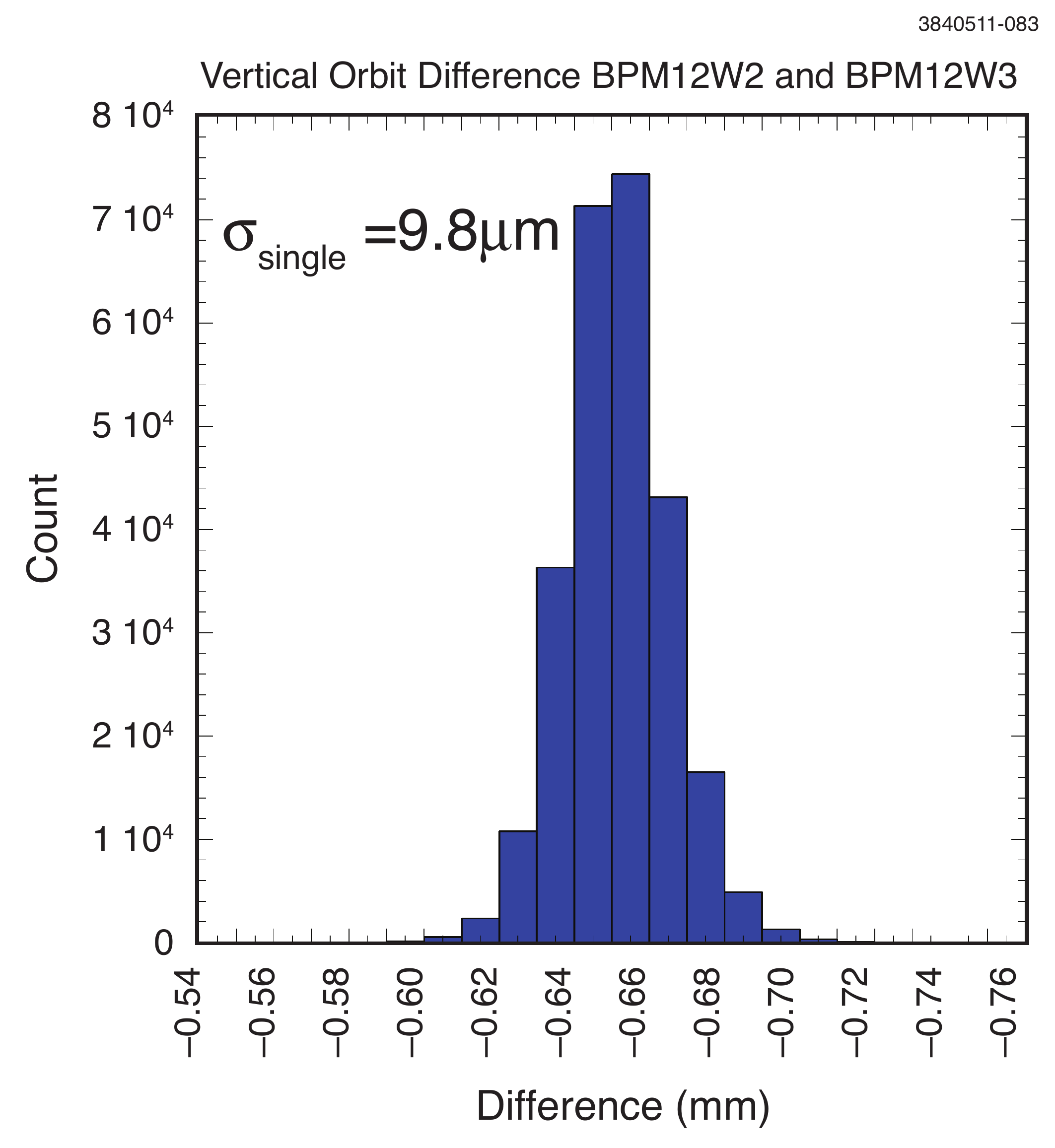}
\caption[Vertical orbit residuals to the fitted trajectories between pairs of closely located detectors at 12W
\cite{IPAC10:MOPE089}.]{\label{fig:Triplet 2} Vertical orbit residuals to the fitted trajectories between pairs of closely located detectors on a straight beam pipe at 12W (between quadrupoles 12W and 13W in CESR, the location of
the CESR diagnostic triplet.)  (The average differences not equalling zero indicate the absolute differences in the centers of the BPMs, which were not calibrated at the time of these measurements).}
\end{figure}

A special diagnostic triplet location has been used to study the resolution
and stability of the system.  The triplet consists of three sets of
detectors mounted in close proximity on a single vacuum chamber.  The
vertical trajectory is fitted turn-by-turn and the residuals to the fits
give the error for the position measurements with beam motion removed.
Figure~\ref{fig:Triplet 2} shows a set of residuals for vertical orbit
differences between pairs of triplet detectors.  The uncertainties
shown in the plots are the computed uncertainties for single
position measurements for a CBPM module, based on these data.  The
histograms include 256K~turns of data (0.67~sec duration) taken
simultaneously with each detector.  The effective resolution
corresponds to the standard deviation of each distribution, which is
consistent with our goal for the system and comparable to the single
shot position resolution, which was estimated above.

\subsection{Measurements of Betatron Phase Advance and Coupling}

The software processing in each CBPM module permits measurements of
betatron phase and coupling.  The basic formalism employed for
extracting betatron phase advance from BPM data is described
elsewhere\cite{PRSTAB3:092801}. Shaker magnets or stripline kickers,
phased locked to a beam position sensor by devices called tune
trackers, are utilized to resonantly excite a single bunch in CESR in the two
normal dipole modes (corresponding to approximately horizontal and
vertical). The phases of the tune tracker drives are digitized synchronously 
with each turn of the bunch circulating in CESR and these 9-bit digital values 
are inserted into the CBPM timing clock. When the CBPM modules are 
triggered to record turn-by-turn positions, they also record the tune tracker 
phase information turn-by-turn. After the data is acquired, the CBPM
modules, utilizing lookup tables to reconstruct the trigonometric functions, 
integrate the position data to project the positions into
the cosine-like and sine-like components with respect to the tune
tracker's phase.  This permits an accurate measurement of the beam's
oscillation phase and amplitude even in the presence of variations
in the betatron tunes. The cosine-like and sine-like components for
each BPM electrode are returned to the CBPM servers for offline
processing and analysis.  Typically 40,960 turns of turn-by-turn
data are acquired and analyzed for the betatron phase measurements
to cover six periods of the 60~Hz mains frequency.

\begin{figure}[h] %  figure placement: here, top, bottom, or page
   \centering
   \includegraphics[height=3in]{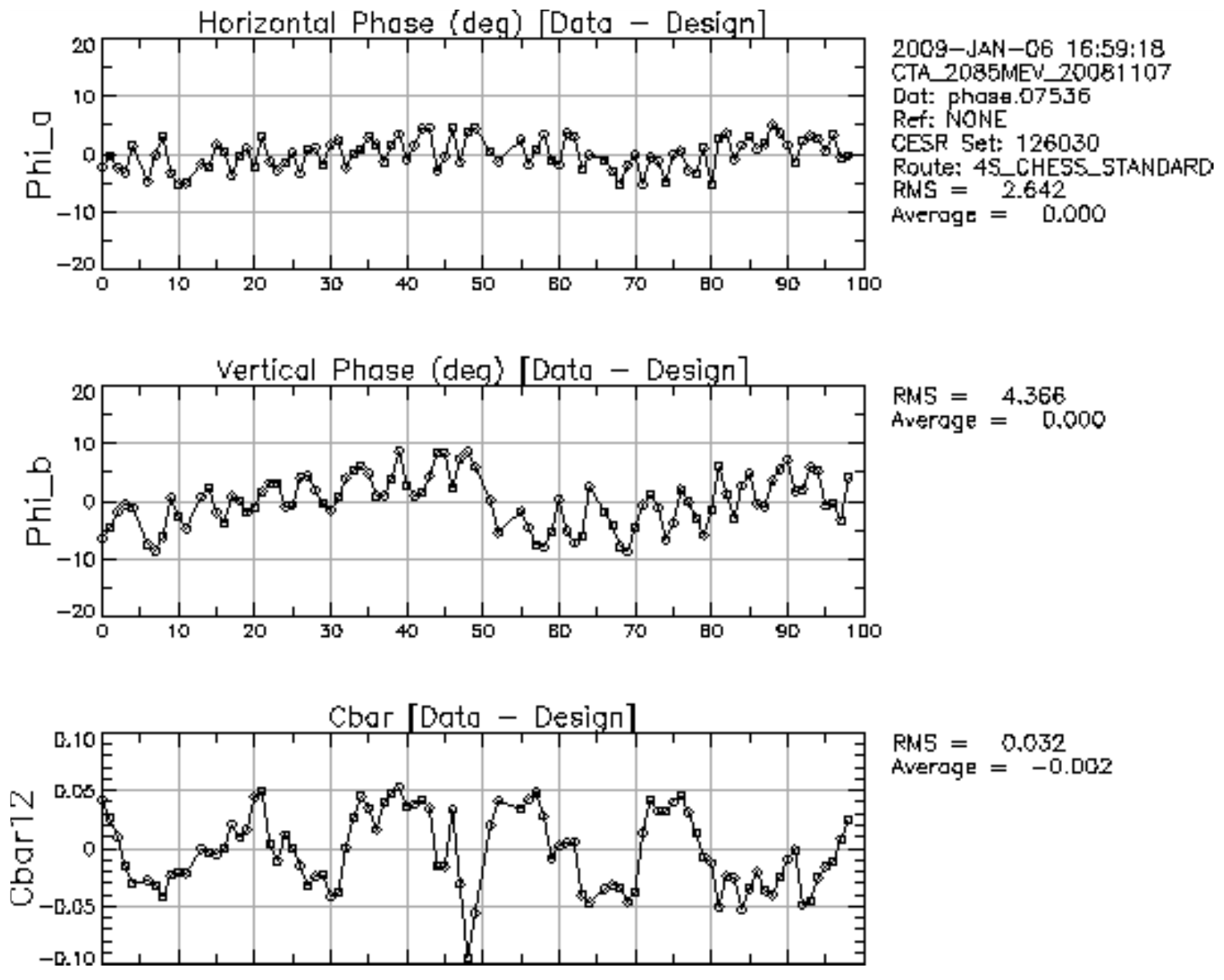}
   \caption[Set of Phase and Coupling Measurements.]{\label{fig:Phase-Coupling} An example of a set of phase and coupling measurements.  The two upper plots are the deviations from design for the betatron phases for the A-mode (horizontal-like) and B-mode (vertical-like) betatron dipole modes.  The lower plot is the off-diagonal element of the $\bar{C}$ horizontal-to-vertical coupling matrix.  The horizontal scales on the plots are BPM number with the data set representing one complete cycle around CESR.  The vertical scales for the phases are in degrees.  The vertical scale for $\bar{C}_{12}$ is unitless.}
\end{figure}

\begin{figure}[h] %  figure placement: here, top, bottom, or page
   \centering
   \includegraphics[height=3in]{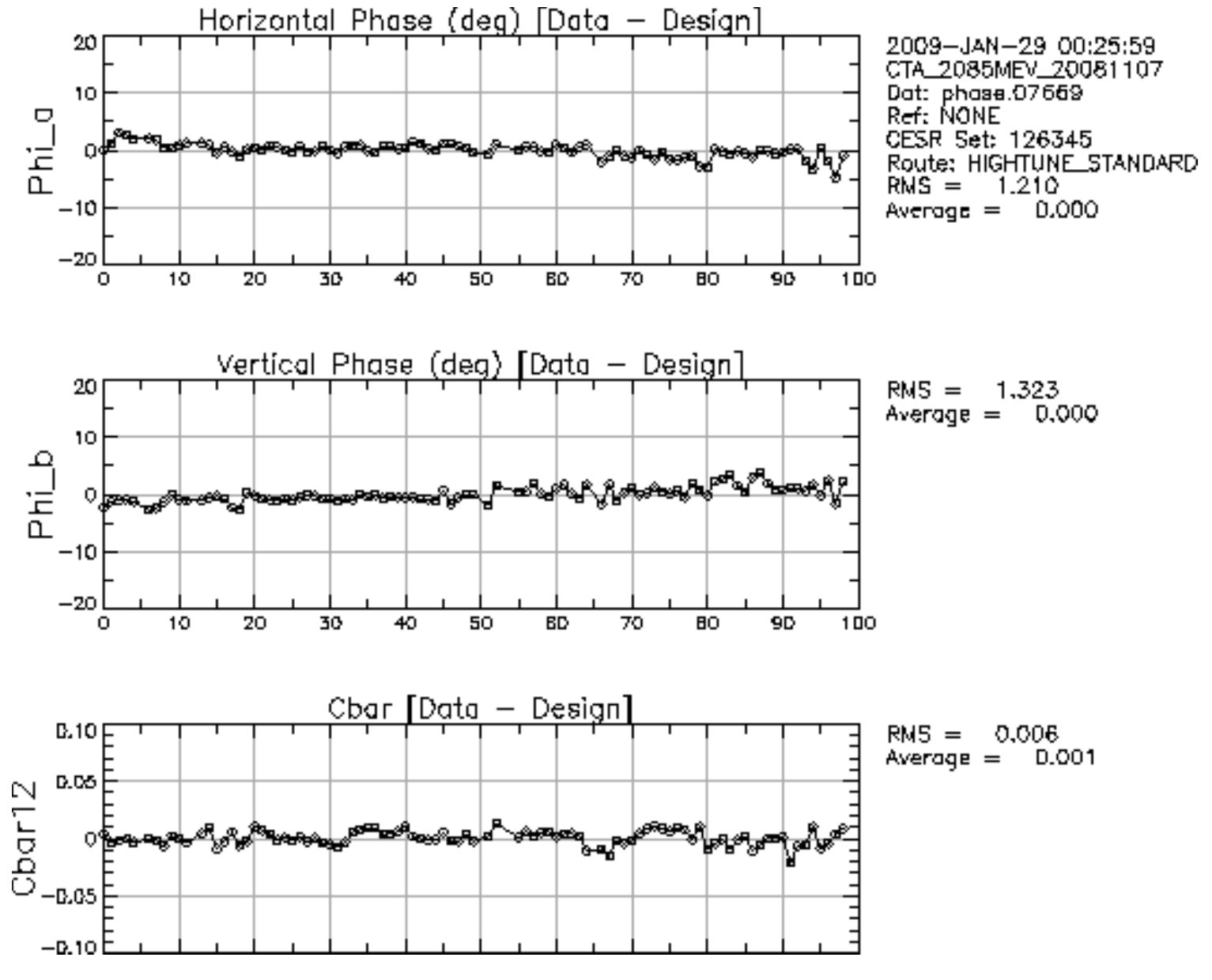}
   \caption[Set of Phase and Coupling Measurements.]{\label{fig:Phase-Coupling-Fixed} After applying phase and coupling correction to the data in the preceding figure.  The two upper plots are the deviations from design for the betatron phases for the A-mode (horizontal-like) and B-mode (vertical-like) betatron dipole modes.  The lower plot is the off-diagonal element of the $\bar{C}$ horizontal-to-vertical coupling matrix.  The horizontal scales on the plots are BPM number with the data set representing one complete cycle around CESR.  The vertical scales for the phases are in degrees.  The vertical scale for $\bar{C}_{12}$ is unitless.}
\end{figure}

An example of a phase and coupling measurement relative to the
optics design is shown in Figure~\ref{fig:Phase-Coupling}.  The
phase measurements display betatron phase waves propagating at twice
the betatron phase advance.  In the vertical, a quadrupole gradient
error is clearly visible as a discontinuity for the average phase
error near BPM detector 50.  Figure~\ref{fig:Phase-Coupling-Fixed}
displays the preceding data after phase and coupling corrections
have been applied.

The uncertainty of the phase and coupling measurements has been
estimated by recording a set of ten phase measurements in
succession. The averages of the phase and $\bar{C}_{12}$ coupling
distributions has been computed detector by detector, resulting in the
average RMS phase uncertainties for the horizontal phase of 0.060
degrees and for the vertical phase of 0.034 degrees and the average
RMS $\bar{C}_{12}$ uncertainty of $6.9\times 10^{-4}$.

%By routing the horizontal tune tracker signal to either the CESR RF
%phase external input or to the longitudinal feedback cavity, an
%energy oscillation may be excited.   Performing a phase measurement
%and extracting the oscillation amplitude produce dispersion
%measurements in both horizontal and vertical directions.  Since the
%tune tracker can excite a single bunch, using a gated input to the
%horizontal and vertical stripline kickers or the longitudinal
%feedback cavity, the betatron phase and dispersion may be measured
%for that bunch alone when it is a member of a multiple bunch fill.

\subsection{Status and Performance}
%\label{sssec:cesr_conversion.beam_instr.bpm.performance}
Presently there are 110~digital BPM readout modules installed in CESR.
Generally 100 of these modules are in routine use for beam diagnostics
and machine studies with the remainder either available for orbit and
trajectory measurements, for bunch-by-bunch current monitoring or for
instrumentation hardware and software diagnostics.  The BPM system has
regularly been used to measure the bunch-by-bunch positions of a train
of 4~ns spaced bunches with a minimum of cross talk between bunch signals.

The hardware and infrastructure of the CESR BPM system has been
functioning since 2009 and has been in active use since its
installation. Although the CBPM system is in routine use at CESR,
development continues for improvements in data storage, analysis and
diagnostic software.  

Some of the development work that is in progress using CBPM module includes 1) the installation
of 3 BPMs in each of the electron and positron transport lines from the Synchrotron 
injector, triggered with each injection cycle, to measure the trajectory of the 
incoming bunches from the Synchrotron and 2) a single BPM in the Synchrotron, triggered at 
each injection cycle to quantify the amplitudes of oscillation for the bunches injected from the 
LINAC.  At the time of this paper's preparation, programming is underway for the Synchrotron 
BPM readout module.

The transfer line BPM readouts have been in routine use during 2016 and their
performance has been under study.  Injection from the LINAC and Synchrotron into
CESR is a top-off process, where the charge per bunch may be 2 to 3 orders of magnitude
lower than the stored beam current.  As a result this requires the use the CBPM's high
gain amplifiers before digitization.  The observations to date are summarized here qualitatively.  
After the modules' amplifier gains are adjusted, the signals from the electron beam are 
typically large enough to have similar digitizer uncertainties as the stored beam.  Even after
applying full module gain amplification, the positron bunch signals are usually well less
than the full scale for the digitizers, implying the digitizing error have a larger influence
on the beam position measurements for the positron beam.  When compared to the stored beam,
the timing jitter from pulse-to-pulse is not expected to be a major problem, since the energy acceptance 
of the Synchrotron limits the acceptible timing jitter of LINAC to be $\pm5$~psec at an absolute maximum
with the variation of timing for the beam extracted after acceleration in the Synchrotron is
expected to be less than this.

The standard time in procedure many injection cycles to generate a timing scan (Figure~\ref{fig:BPM_SIgnalSampling})
Due to pulse-to-pulse intensity variations of the Synchrotron beam, it is difficult to achieve 
as precise timing for each button as for CESR, using the standard timing algorithm.  The
measured variation of the electron beam position has uncertainties of $\pm0.32$~mm and 
$\pm0.34$~mm in the horizontal and vertical direction, respectively.  Most of the position 
uncertainties during operations are due to pulse-to-pulse positioning errors when the LINAC
bunches are injected into the Synchrotron, which result in pulse-to-pulse variations in the
position of Synchrotron bunches during extraction for injection into CESR.

\nopagebreak

\section{Summary}
\label{sec:summary}

This paper has described the upgrade of the CESR beam position
monitoring instrumentation, which has been developed for use in the
\cesrta~ program for the investigation of storage ring beam
dynamics.  In particular the new CBPM modules have achieved the
design goals set for the \cesrta~  program.  This has been an
important tool for the \cesrta~ studies focusing on the methods for
low emittance tuning of the beam\cite{PRSTAB17:044003}, on the causes of intra-beam
scattering of single bunches\cite{PRSTAB18:064402} and on the production and interaction
of bunches within trains with electron clouds, which have been
produced by photo-electrons from synchrotron radiation and secondary
emission.

\bibliographystyle{JHEP}
\bibliography{CesrTA}

\providecommand{\href}[2]{#2}\begingroup\raggedright\begin{thebibliography}{10}

\bibitem{JINST10:P07012}
M.~Billing, \emph{{T}he conversion of {CESR} to operate as the {T}est
  {A}ccelerator, {C}esr{TA}. {P}art 1: overview},
  \href{http://dx.doi.org/10.1088/1748-0221/10/07/P07012}{\emph{J. Instrum.}
  {\bfseries 10} (July, 2015) }.

\bibitem{JINST10:P07013}
M.~G. Billing and Y.~Li, \emph{The conversion of {CESR} to operate as the test
  accelerator, {CesrTA}, part 2: Vacuum modifications},
  \href{http://dx.doi.org/10.1088/1748-0221/10/07/P07013}{\emph{J. Instrum.}
  {\bfseries 10} (July, 2015) }.

\bibitem{JINST11:P04025}
M.~G. Billing, J.~V. Conway, J.~A. Crittenden, S.~Greenwald, Y.~Li, R.~E.
  Meller et~al., \emph{The conversion of {CESR} to operate as the test
  accelerator, {CesrTA}, part 3: Electron cloud diagnostics},
  \href{http://dx.doi.org/10.1088/1748-0221/11/04/P04025}{\emph{J. Instrum.}
  {\bfseries 11} (Apr., 2016) }.

\bibitem{PRSTAB8:062802}
R.~W. Helms and G.~H. Hoffstaetter, \emph{Orbit and optics improvement by
  evaluating the nonlinear beam position monitor response in the {C}ornell
  {E}lectron {S}torage {R}ing},
  \href{http://dx.doi.org/10.1103/PhysRevSTAB.8.062802}{\emph{Phys. Rev. ST
  Accel. Beams} {\bfseries 8} (June, 2005) }.

\bibitem{PRSTAB13:092802}
D.~L. Rubin, M.~Billing, R.~Meller, M.~Palmer, M.~Rendina, N.~Rider et~al.,
  \emph{Beam based measurement of beam position monitor electrode gains},
  \href{http://dx.doi.org/10.1103/PhysRevSTAB.13.092802}{\emph{Phys. Rev. ST
  Accel. Beams} {\bfseries 13} (Sept., 2010) }.

\bibitem{PRSTAB14:072804}
A.~Wolski, D.~Rubin, D.~Sagan and J.~Shanks, \emph{Low-emittance tuning of
  storage rings using normal mode beam position monitor calibration},
  \href{http://dx.doi.org/10.1103/PhysRevSTAB.14.072804}{\emph{Phys. Rev. ST
  Accel. Beams} {\bfseries 14} (July, 2011) }.

\bibitem{PAC01:TPAH064}
M.~Palmer, J.~Dobbins, D.~Hartill and C.~Strohman, \emph{An upgrade for the
  beam position monitoring system at the {C}ornell {E}lectron {S}torage
  {R}ing},  in \emph{Proceedings of the 2001 Particle Accelerator Conference,
  Chicago, IL} (P.~Lucas and S.~Webber, eds.), pp.~1360--1362, IEEE, 2001.

\bibitem{PAC03:WPAG004}
J.~C. Smith, M.~A. Palmer, D.~L. Rubin and D.~C. Sagan, \emph{Diagnosis of
  optical errors with a precision {BPM} system at {CESR}},  in
  \emph{Proceedings of the 2003 Particle Accelerator Conference, Portland, OR}
  (J.~Chew, P.~Lucas and S.~Webber, eds.), pp.~2267--2269, IEEE, 2003.

\bibitem{PAC03:WPPB027}
M.~A. Palmer, J.~A. Dobbins, B.~Y. Rock, C.~R. Strohman and J.~R. Moffitt,
  \emph{System level implementation of beam position monitors with local data
  processing capability for the {C}ornell {E}lectron {S}torage {R}ing},  in
  \emph{Proceedings of the 2003 Particle Accelerator Conference, Portland, OR}
  (J.~Chew, P.~Lucas and S.~Webber, eds.), pp.~2473--2475, IEEE, 2003.

\bibitem{PAC05:TPPP011}
M.~G. Billing, J.~A. Crittenden and M.~A. Palmer, \emph{Investigations of
  injection orbits at {CESR} based on turn-by-turn {BPM} measurements},  in
  \emph{Proceedings of the 2005 Particle Accelerator Conference, Knoxville, TN}
  (C.~Horak, ed.), pp.~1228--1230, IEEE, 2005.

\bibitem{IPAC10:MOPE089}
M.~A. Palmer, M.~G. Billing, R.~E. Meller, M.~C. Rendina, N.~T. Rider, D.~L.
  Rubin et~al., \emph{{CESR} beam position monitor system upgrade for {CesrTA}
  and {CHESS} operations},  in \emph{Proceedings of the 2010 International
  Particle Accelerator Conference, Kyoto, Japan}, pp.~1191--1193, ACFA, 2010.

\bibitem{IPAC11:TUPC052}
A.~Wolski, D.~Rubin, D.~Sagan and J.~Shanks, \emph{Normal mode {BPM}
  calibration for ultralow emittance tuning in lepton storage rings},  in
  \emph{Proceedings of the 2011 International Particle Accelerator Conference,
  San {Sebasti\'{a}n}, Spain}, pp.~1114--1116, EPS-AG, 2011.

\bibitem{CLNS:12:2084}
\emph{The {CESR} {T}est {A}ccelerator electron cloud research program: Phase
  {I} {R}eport},  Tech. Rep. CLNS-12-2084, {LEPP}, Cornell University, Ithaca,
  NY, Jan., 2013.

\bibitem{IEEENS26:4078to4079}
R.~Helmke, D.~Rice and S.~Ball, \emph{{Interface Hardware for the CESR Control
  System}}, \href{http://dx.doi.org/10.1109/TNS.1979.4330018}{\emph{IEEE Trans.
  Nuclear Science} {\bfseries 26} (June, 1979) }.

\bibitem{IPAC12:THPPR015}
M.~J. Forster, S.~Ball, L.~Bartnik, D.~Bougie, R.~Helmke, M.~Palmer et~al.,
  \emph{{CESR} control system upgrade to linux high availability cluster},  in
  \emph{Proceedings of the 2012 International Particle Accelerator Conference,
  New Orleans, LA}, pp.~3999--4001, IEEE, 2012.

\bibitem{Strohman:CBINET}
C.~Strohman and T.~Wilksen, \emph{{C}{B}{I}{\_}{N}{E}{T} reference},  tech.
  rep., Cornell~University, 2006.

\bibitem{BIW92:55to77}
M.~Billing, \emph{Introduction to beam diagnostics and instrumentation for
  circular accelerators},  in \emph{Proceedings of the 1992 Beam
  Instrumentation Workshop}, pp.~55--77, 1992.
\newblock \href{http://dx.doi.org/10.1063/1.44330}{DOI}.

\bibitem{Billing:UNDBPM}
M.~Billing, \emph{Unpublished calculation},  tech. rep., Cornell University,
  2012.

\bibitem{PRSTAB3:092801}
D.~Sagan, R.~Meller, R.~Littauer and D.~Rubin, \emph{Betatron phase and
  coupling measurements at the {C}ornell {E}lectron/positron {S}torage {R}ing},
  \href{http://dx.doi.org/10.1103/PhysRevSTAB.3.092801}{\emph{Phys. Rev. ST
  Accel. Beams} {\bfseries 3} (Sept., 2000) }.

\bibitem{PRSTAB17:044003}
J.~Shanks, D.~L. Rubin and D.~Sagan, \emph{Low-emittance tuning at the
  {C}ornell {E}lectron {S}torage {R}ing {T}est {A}ccelerator},
  \href{http://dx.doi.org/10.1103/PhysRevSTAB.17.044003}{\emph{Phys. Rev. ST
  Accel. Beams} {\bfseries 17} (Apr., 2014) }.

\bibitem{PRSTAB18:064402}
A.~Chatterjee, K.~Blaser, W.~Hartung, D.~Rubin and S.~Wang, \emph{{Fast ion
  instability at the Cornell Electron Storage Ring Test Accelerator}},
  \href{http://dx.doi.org/10.1103/PhysRevSTAB.18.064402}{\emph{Phys. Rev. ST
  Accel. Beams} {\bfseries 18} (June, 2015) }.

\end{thebibliography}\endgroup

\end{document}